\documentclass[useAMS,usenatbib]{mn2e}
\usepackage{graphicx}
\usepackage{epstopdf}
\usepackage{amsmath}
\usepackage{times}
\usepackage{url}

\title[New method for estimation of asteroid coordinates]{A new method based on the subpixel Gaussian model for accurate estimation of asteroid coordinates}

\author[Savanevych et al.]
{Savanevych, V. E.$^1$\thanks{vadym@savanevych.com}, Briukhovetskyi, O. B.$^2$, Sokovikova, N. S.$^1$, Bezkrovny, M. M.$^3$, \newauthor Vavilova, I. B.$^4$, Ivashchenko, Yu. M.$^{4,5}$, Elenin, L. V.$^6$, Khlamov, S. V.$^1$, Movsesian, Ia. S.$^1$, \newauthor Dashkova, A. M.$^3$, Pogorelov A.V.$^1$\\
$^1$Kharkiv National University of Radio Electronics, 14 Lenin ave., Kharkiv 61166, Ukraine\\
$^2$Kharkiv Representation of the General Customer's Office of the State Space Agency of Ukraine, 1 Akademika Proskury St., Kharkiv 61070, Ukraine\\
$^3$Zaporizhya Institute of Economics and Information Technologies, 16-B Kiyashka St., Zaporizhya 69015, Ukraine\\
$^4$Main Astronomical Observatory of the NAS of Ukraine, 27 Akademika Zabolotnogo St., Kyiv 03680, Ukraine\\
$^5$Andrushivka Astronomical Observatory, Galchyn, Zhytomyr Region 13400, Ukraine\\
$^6$Keldysh Institute of Applied Mathematics of the RAS, 4 Miusskaya sq., Moscow 125047, Russian Federation\\}

\pagerange{\pageref{firstpage}--\pageref{lastpage}}\pubyear{2014}

\begin{document}
\label{firstpage}
\maketitle

\begin{abstract}
We describe a new iteration method to estimate asteroid coordinates, which is based on the subpixel Gaussian model of a discrete object image. The method operates by continuous parameters (asteroid coordinates) in a discrete observational space (the set of pixels potential) of the CCD frame. In this model, a kind of the coordinate distribution of the photons hitting a pixel of the CCD frame is known a priori, while the associated parameters are determined from a real digital object image.
The developed method, being more flexible in adapting to any form of the object image, has a high measurement accuracy along with a low calculating complexity due to a maximum likelihood procedure, which is implemented to obtain the best fit instead of a least-squares method and Levenberg-Marquardt algorithm for the minimisation of the quadratic form.

Since 2010, the method was tested as the basis of our CoLiTec (Collection Light Technology) software, which has been installed at several observatories of the world with the aim of automatic discoveries of asteroids and comets on a set of CCD frames. As the result, four comets (C/2010 X1 (Elenin), P/2011 NO1(Elenin), C/2012 S1 (ISON), and P/2013 V3 (Nevski)) as well as more than 1500 small Solar System bodies (including five NEOs, 21 Trojan asteroids of Jupiter, and one Centaur object) were discovered. We discuss these results that allowed us to compare the accuracy parameters of a new method and confirm its efficiency.

In 2014, the CoLiTec software was recommended to all members of the Gaia-FUN-SSO network for analysing observations as a tool to detect faint moving objects in frames.

\end{abstract}

\begin{keywords}
methods: data analysis; minor planets, asteroids, comets; techniques: image processing.
\end{keywords}

\section{Introduction}
There are many methods for determining an asteroid position during observations with a CCD camera. For example, the Full Width at Half Magnitude (FWHM) approach \citep{Gary}, which is based on the analytical description of the object images on the CCD frame, as well as other methods, in which the position of an object's maximum brightness on a CCD image is taken as its coordinates (\citet{Miura}).

Most of these methods have a common feature. They use PSF-fitting (Point-Spread Function) to approximate the object image and get the information about regularities in the distribution of the registered photons on CCD frame(see, for detail, \citet{Yanagisawa}, \citet{Gural}, \citet{Veres}, \citet{DellOro}, \citet{Lafreniere}, \citet{Zacharias}),
Among the models of photons distribution, which are used more often, we note
the two-dimensional Gaussian model (\citet{Veres}, \citet{Zacharias}, \citet{Babu}), Moffat model (\citet{Bauer}, \citet{Izmailov}) or Lorentz model (\citet{Zacharias}, \citet{Izmailov}).
These models are usually described by the continuous functions, while the CCD images are discrete ones.
Such an approach was criticised reasonably by \citet{Bauer}. The principal disadvantage is that these models work well only with a large amount of the data. It leads to the fact that, firstly, the computation process becomes much more complicated, and, secondly, the problem related with the adequacy of the used estimations of PSF parameters cannot be solved. As a result, the error of the coordinate determination of the observed celestial objects has to be increasing.

In addition to the aforementioned disadvantage,
the existing methods do not pay sufficient attention for taking the noise component of the object image into account.
It is assumed that its registration and compensation are performed during the preliminary stage of image processing (\citet{Gural}) or that the object image is exempted from noise according to the accepted signal-to-noise model (\citet{Lafreniere}, \citet{Izmailov}). At the same time, the error introduced by the operation of removing the noise component from the object image is not considered in the subsequent procedure of coordinate determination.

The errors of CCD astrometry are traditionally divided into the instrument, reduction, reference catalogue and measurement errors.

The first ones traditionally include errors of instrumental parameters as shutter delay and clock correction, which result in incorrect timing. The second error category (reduction) is associated with the method related to the standard and measured coordinates and depends on the choice of the algorithm relation between these coordinates. For example, being not sufficient for the wide-angle astrographs such an error depends strongly on the choice of the type and degree of the polynomial approximation in Turner method. It is important to note that a systematic error of timing is not shared with the coordinate error along the tracking an object in the sky, so it must be caught clearly with a reliable shutter sync.

The reference catalogue errors are divided into three main classes: zonal errors (systematic errors of the reference catalogue); coordinate errors of reference stars in the catalogue epoch; proper motion errors of reference stars. Therefore, a choice of the reference catalogue is very important.
For example, Hipparcos or Tycho catalogues had no errors in the epoch of 1990.0, because an intrinsic accuracy was at the level of millisecond of arc, but for the present epoch there is a necessity to take into account the proper motion errors of reference stars. The solution of this problem, i.e., creation of new huge database of proper motions of stars, is one of the task of the GAIA mission. It is worth noting that the choice of reference catalogue is not so important, when monitoring observations of sky are conducted with the aim of discovery of new Solar system small bodies, because an intrinsic accuracy of a catalogue should not be necessarily maximised as compared with those for the following tracking the discovered object.

The measurement error is related, first of all, to the determination of coordinates of the image centre or the accuracy of a digital approximation of the CCD image (fitting). The attempt to improve the fitting may not lead to expected results if the reference catalogue errors and timing are not  taken attentively into account.

Each of the aforementioned factors could be the source of both systematic and random errors.
Any attempt to reduce one of these errors is impossible without control of other factors. Therefore, the task of the observers is to be responsible for monitoring all the possible sources of errors. The aim of this paper is to help the observers to refine the coordinate measurements of the object image on the CCD frame and to control the errors of the measured coordinates.

With this aim, we developed a new method for accurately estimating asteroid coordinates on a set of CCD frames, which is based on the subpixel Gaussian model of the discrete image of an object.
In this model, a kind of the coordinate distribution of the photons hitting a pixel on the CCD frame is known a priori, while its parameters can be determined from the real digital image of the object. Our method has low computational complexity due to the use of equations of maximum likelihood, as well as the proposed model is more flexible, adapting to any form of the real image of the object.
For example, in fact PSF is a super-pixel function because it describes the changes in the brightness of a pixels of celestial object image. We propose to use the density function of coordinates of hitting photons from celestial object instead of the PSF. This function is sub-pixel one. To get the Point-Spread Function from this function, it should be integrated over the area of determination of each image pixel of object or compact group of objects. It turns out that sub-pixel models are more flexible and may describe the real image more adequately. This effect does not occur in case of bright objects. But applying a more flexible model for the faint objects, we are able to improve the accuracy of the measurements (for example, as comparing with Astrometrica) by 30-50\% (more details are given in the discussion). Our method also takes into account the principal peculiarities of the object image formation on the CCD frame along with the possible irregular distribution of the residual noise component both on the object image and in its vicinity.

\begin{figure*}
	\begin{center}
		\includegraphics[scale=0.25]{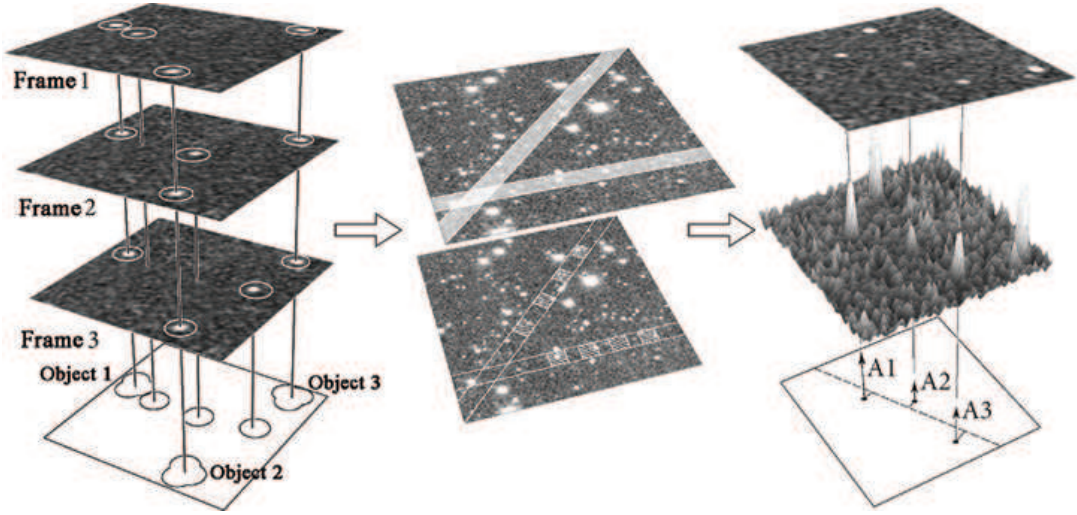}
		\caption{Main steps of object image processing in the CoLiTec software: a) exclusion of stationary objects; b) detection of moving objects; c) analysis of moving objects, where A1, A2, and A3 outline coordinate deviations of moving objects from their trajectory.}
		\label{fig:CoLiTec}
	\end{center}
\end{figure*}

Some generalisations of the proposed method are presented in our previous works (\citet{Sav99}, \citet{Sav06}, \citet{Sav10}, \citet{Sav11} \citet{VavKNIT}, \citet{Sav12}, \citet{VavKPCB}, \citet{VavBA}, \citet{Sav14}). Since 2010, due to its application with the CoLiTec (Collection Light Technology) software, which was installed at several observatories of the world, four comets (C/2010 X1 (Elenin), P/2011 NO1(Elenin), C/2012 S1 (ISON), and P/2013 V3 (Nevski)) and more than 1500 small Solar System bodies (including five NEOs, 21 Trojan asteroids of Jupiter, and one Centaur object) were discovered. These results confirm the efficiency of the proposed method. The main stages of image processing with the CoLiTec software are presented in Fig.~\ref{fig:CoLiTec}.

The structure of this article is as follows: we describe a problem statement and the method in Chapters 2-3, respectively. The results based on the testing of this method with the CoLiTec software are presented in Chapter 4.  We compare and discuss the accuracy and other parameters for determining the position of the faint celestial object on the CCD frame obtained by the proposed method and others in Chapter 5. Conclusive remarks are given in Chapter 6.

\section{Problem statement}
If the exposure time is small, the shift of asteroid position in the sky can be ignored. In this case, the asteroid and field stars are imaged as blur spots rather than points on the CCD frame. It is postulated that the coordinates of the signal photons from asteroids and stars hitting the CCD frame have a circular normal distribution with mathematical expectations $x_{t}$, $y_{t}$ and mean-square error (MSE) $\sigma_{ph}$.

It is supposed that a preliminary detection of asteroid has already been conducted before the determination of its coordinates. The result of this detection is a preliminary estimation of the asteroid position on the CCD frame, namely, the determination of coordinates of the pixel, which corresponds to the maximum brightness peak on the asteroid image. We name the set of pixels around this pixel as the area of intraframe processing (AIFP). Thus, the AIFP size ($N_{IPS}$, in pixels) is much larger than the image of the asteroid.

The original CCD image of the celestial object contains harmful interferences such as the read noise, dark currents, irregularity in the pixel-to-pixel sensitivity, sky background radiation, etc. (\citet{Faraji}, \citet{Harris}). Hence, the CCD frame, can be represented as an additive mix of the images of celestial objects and a component, which is formed by this generalised interference. Within the scope of the whole CCD frame
the interference component has a complex structure. However, in a small vicinity of the studied asteroid image, such interference component can be described with a good accuracy as a plane with an arbitrary slope. Such a representation describes well the interference component especially if there is a bright object near the vicinity of the studied AIFP.

The output signals of the CCD matrix pixels $N_{IPS}$ are easily reduced to the relative frequencies $\nu^{*}_{ikt}$ of the photons hitting the $ik^{th}$ pixel on the $t^{th}$ frame:

\begin{equation}\label{eq1}
	\nu^{*}_{ikt}=\frac{A_{ikt}}{\sum\limits_{i,k}^{N_{IPS}}A_{ikt}},
\end{equation}

where $A_{ikt}$ is the brightness of the $ik$ pixel of the CCD matrix.

Then, the result of the observation is the set $\tilde{U}=\nu^{*}_{11t},...,\nu^{*}_{ikt},...,\nu^{*}_{N_{IPSt}}$ of the relative frequencies, which are independent of each other. The theoretical analogues of the measured relative frequencies are the probabilities $\nu_{ikt}(\Theta)$
that during the exposure time the photons hit the $ik^{th}$-pixel of the CCD matrix with the borders $x_{begi}$, $x_{endi}$ in the coordinate $x$ and $y_{begk}$, $y_{endk}$ in the coordinate $y$ on the $t^{th}$ frame.
It is supposed that the angular sizes of the pixel, $\Delta x$ and $\Delta y$, are the same in both coordinates $x$ and $y$.

Thus, the problem statement is as follows: it is needed to develop a method of maximum likelihood estimation of asteroid coordinates on the $t^{th}$ CCD frame using the set of relative frequencies $\nu^{*}_{ikt}$.
It is believed that the likelihood function is differentiable in the vicinity of its global maximum, and its initial approximation is also in the same vicinity. The set of the estimated parameters $\Theta$ includes the asteroid coordinates $x_{t}$ and $y_{t}$ on the $t^{th}$ frame and the mean square error of the coordinates of photons hitting the CCD matrix, $\sigma_{ph}$.

To introduce a new method, we use two functions. The density distribution of a normally distributed random variable $z$ with mathematical expectation $m_{z}$ and dispersion $\sigma^2$ is determined by the expression:

\begin{equation}\label{eq2}
	N_{z}(m_{z},\sigma^{2})=\frac{1}{\sqrt{2\pi\sigma}}exp(-\frac{1}{2\sigma^{2}}(z-m_{z})^{2}).
\end{equation}

The probability $F_{z}$ that a random variable $z$ is within the closed interval $[z_{beg}, z_{end}]$ is:

\begin{equation}\label{eq3}
	F_{zi}(m_{z},\sigma^{2})=\int_{z_{end}}^{z_{beg}}N_{z}(m_{z};\sigma^{2})dz.
\end{equation}

\section{Task solution}
For achieving the maximum accuracy of the estimations of the object position on the frame, the discretisation factor needs to be taken into account, because we should estimate the continuous parameters (coordinates of objects) at the discrete set of the measured values (the brightness of the CCD matrix pixels). The general view of the maximum likelihood estimation of the object position can be expressed by:

\begin{equation}\label{eq4}
	\sum_{i,k}^{N_{IPS}}\frac{\nu^{*}_{ikt}}{\nu_{ikt}(\Theta)}\frac{\partial\nu_{ikt}(\Theta)}{\partial\Theta_{m}}=0,
\end{equation}

where $\Theta$ is the set of the estimated parameters $x_{t}$, $y_{t}$, and $\sigma_{ph}$.

The relation between the probability $\nu_{ikt}(\Theta)$  that photons hit the $ik^{th}$ pixel (Eq. 4)  and the function of coordinates distribution  $f(x,y)$  of the incidence of photons from the object on the CCD matrix has the form:

\begin{equation}\label{eq5}
	\nu_{ikt}(\Theta)=\int_{x_{beg i}}^{x_{end{i}}}\int_{y_{beg k}}^{y_{end k}}f(x,y)dxdy.
\end{equation}

After the compensation of the noise component on the CCD image, the function $f(x,y)$ could be expressed as the weighted mix of normal and uniform probability distributions:

\begin{equation}\label{eq6}
	f(x,y,\Theta)=p_{0}+\frac{p_{1}}{2\pi\sigma^{2}_{ph}}\nonumber\\
\end{equation}
\begin{equation}\label{eq6}
	\exp\{-\frac{1}{2\sigma^{2}_{ph}}[(x-x_{t})^{2}+(y-y_{t})^{2}]\},
\end{equation}

where $p_{1}=1-p_{0}$ is the relative weight of the signal photons of the object;
$p_{0}$ ($0\leq p_{0}<1$) is the relative weight of the residual noise photons of the CCD matrix after the compensation of the flat generalised interference; $x_{t}$ and $y_{t}$ are the object coordinates on the $t^{th}$ frame at the time $t^{t}$ corresponding to the mathematical expectations of the coordinates of incidence of the signal photons.

The probability (Eq.5) that photons hit the CCD matrix pixels can be written as follows:

\begin{equation}\label{eq7}
	\nu_{ikt}(\Theta)=I_{iktnoise}+I_{ikts},
\end{equation}

where $I_{ikts}=p_{i}F_{xi}(xi; \sigma_{phi}^{2}) F_{yk}(y_{t},\sigma_{ph}^{2})$
is the probability that signal photons $I_{ikt}$ hit the $ik^{th}$ pixel of the CCD matrix;
and $I_{iktnoise}=\Delta_{CCDp0}^{2}$ is the probability that the noise residuary photons hit the $ik^{th}$ pixel of the CCD matrix; $\Delta_{CCD}=\Delta_{x}= \Delta_{y}$.

The derivative from the probability (Eq. 7) in the $x$ coordinate is determined by the expression:

\begin{equation}\label{eq8}
	\frac{d\nu_{ikt}(\Theta)}{dx_{t}}=\frac{p_{1}F_{yk}(y_{t};\sigma_{ph}^{2})F_{xi}(x_{t};\sigma_{ph}^{2})}{\sigma_{ph}}(m_{xi}^{loc} - x_{t}),
\end{equation}

where

\begin{eqnarray}\label{eq9}
	m_{xi}^{loc}=m_{x}+\frac{\sigma^{2}}{F_{xi}(m_{x};\sigma^{2})}(N_{xendi}(m_{x};\sigma^{2})-N_{xbegi}(m_{x};\sigma^{2}))\nonumber
\end{eqnarray}

is the local (on the closed interval [$xbegi; xendi$]) mathematical expectation of the normally distributed random value $x$ with a mathematic expectation $m_{x}$ and dispersion $\sigma^{2}$. The derivative from the probability (Eq. 7) in the $y$ coordinate has the same expression as Eq. 8.

 The system of equations of the maximum likelihood for the studied AIFP pixels in the case while the asteroid position is estimated only, will take the form:

\begin{equation}\label{eq9}
	\begin{cases}
		\hat{x_{t}}=\frac{\sum_{i,k}^{N_{IPS_{s}}}\nu^{*}_{ikt}\lambda_{ikt}m_{xi}^{loc}}{\sum_{i,k}^{N_{IPS_{s}}}\nu^{*}_{ikt}\lambda_{ikt}};\\
		\hat{y_{t}}=\frac{\sum_{i,k}^{N_{IPS_{s}}}\nu^{*}_{ikt}\lambda_{ikt}m_{yk}^{loc}}{\sum_{i,k}^{N_{IPS_{s}}}\nu^{*}_{ikt}\lambda_{ikt}},
	\end{cases}
\end{equation}

where $N_{IPS_{s}}$ is the quantity of pixels in the part of AIFP area (the area, where the signal from the object is expected); $\hat{x_{t}}$ and $\hat{y_{t}}$ are the estimations of asteroid coordinates; and $\lambda_{ikt}$ is the part of photons from the celestial object in the $ik^{th}$ pixel of the $t^{th}$ CCD frame. The latter value is determined by:

\begin{equation}\label{eq10}
	\lambda_{ikt}=\frac{p_{1}F_{yk}(y_{t};\sigma_{ph}^{2})F_{xi}((x_{t};\sigma_{ph}^{2})}{\nu_{ikt}(\Theta)}.
\end{equation}

To estimate the MSE of the coordinates of the photons hitting the CCD frame from the asteroid, we use the equation of the maximal likelihood:

\begin{equation}\label{eq11}
	{\hat\sigma_{ph}^{2}}=\frac{\sum_{i,k}^{N_{IPSs}}\nu_{ikt}^{*}\lambda_{ikt}((m_{xi}^{loc}-\hat{x_{t}})^2+(m_{yk}^{loc}-\hat{y_{t}})^2)}{2\sum_{i,k}^{N_{IPSs}}\nu_{ikt}^{*}\lambda_{ikt}}.
\end{equation}

We cannot exclude completely the generalised noise interference. By this reason, to take into account the relative weight of the signal photons, we use a standard estimation of the weights of the discrete mix of the probability distributions (\citet{Lo}):

\begin{equation}\label{eq12}
	\hat{p_{1}}=\frac{1}{N_{IPSs}}\sum_{i,k}^{N_{IPSs}}\lambda_{ikt};\hat{p_{0}}=1-\hat{p_{1}}.
\end{equation}

Therefore, the local mathematical expectation of coordinates of the object position is a function of the relevant coordinates, and Eq. 9 gives a system of transcendental equations that can be solved by the method of successive approximations (see, e.g., \citet{Burden}).

The algorithm of the estimation of object coordinates consists of two successive operations.

The first operation is to split the statistics of the AIFP pixels into the statistics of the signal and statistics of the residual interference. It is performed for pixels from the AIFP area where the object is expected.
According to the $\Theta$ values calculated from the previous iteration, the photons of pixel are divided into those belonging to the object and to the residual interference. The photons belonging to the object are analysed to determine estimation of its position. Thus, coordinates of the local maximum in the object image, around which the AIFP area is formed, are used as the initial approximation. The result of this operation is a set of split coefficients $\lambda_{ikt}$.

The second operation provides estimation of the object coordinates based on the statistics, which are obtained during the operation of splitting. It is conducted in a strongly determined way from Eq. 9 to Eq. 12.
The result $\hat\Theta_{n}$ of this operation serves as an initial approximation for the operation of splitting at the next iteration step. The iteration process is continued until the difference between  $\hat\Theta_{n}$ and $\hat\Theta_{n-1}$ becomes smaller than the predetermined value, for example, 0.1\% of the angle size of a pixel.

The analysis of the iteration process shows that its convergence is provided while the following conditions are fulfilled:

\begin{equation}\label{eq13}
	dx=\frac{|x_{0}-x_{true}|}{\sigma_{x}}<6,dy=\frac{|y_{0}-y_{true}|}{\sigma_{y}}<6,
\end{equation}

where $dx$ and $dy$ are the relative distances between the initial and actual positions of the object; $x_{0}$ and $y_{0}$ are the initial approximation of the object coordinates, $x_{true}$ and $y_{true}$ are the actual object coordinates; and $\sigma_{x}$ and $\sigma_{y}$ are the MSEs of coordinates of the signal photons hitting the CCD matrix.

The observations based on the proposed method have shown that  this condition is almost always fulfilled for the real images of asteroids and stars, and, in most cases, the relative distance is not more than $1.0\div15$.

The opportunity to divide the AIFP area into the interference area (pixels that have registered photons only from interferences) and the object area (pixels that have registered photons from the object and interference) yields a more simple and reliable algorithm of estimation of the flat interference component (as compared with the estimation in the common system of maximum likelihood equations). Namely, there is an independent estimation by the method of least squares (MLS). Thus, the density of the coordinate distribution of the photons from the residual interference will represent an equation of the plane with an arbitrary slope:

\begin{equation}\label{eq14}
	f_{noise}(x,y)=A_{noise}x+B_{noise}y+C_{noise}.
\end{equation}

The probability that these photons will hit the $ik^{th}$ pixel can be given by analogy with Eq. 5:

\begin{equation}\label{eq15}
	\nu_{iktnoise}^{*}(\Theta_{noise})=A_{noise}^{int}x_{ik}+B_{noise}^{int}y_{ik}+C_{noise}^{int},
\end{equation}

where $\nu_{iktnoise}^{*}$ is the measured frequency of the noise photons hitting the $ik^{th}$ pixel of the CCD matrix;
$A_{noise}^{int}=\Delta_{CCD}^{2}A_{noise}$, $B_{noise}^{int}=\Delta_{CCD}^{2}B_{noise}$, $C_{noise}^{int}=\Delta_{CCD}^{2}C_{noise}$, $\Theta_{noise}^{T}=(A_{noise}^{int},B_{noise}^{int},C_{noise}^{int})$
are the integral parameters of the flat noise component and its vectors;  $x_{ikt}=\frac{x_{endi}+x_{begi}}{2}$, $y_{ikt}=\frac{y_{endk}+y_{begk}}{2}$ are the average values of coordinates of the $ik^{th}$ pixel.

Thus, the probabilities that noise photons will hit the pixels of the studied AIFP depend linearly on the angle coordinates of the centres $x_{j}$ and $y_{j}$ of these pixels and represent, by themselves, a plane with the integral parameters $A_{noise}^{int}$, $B_{noise}^{int}$ and $C_{noise}^{int}$. It is worth noting that the pixels containing the supposed object image should be eliminated before the determination of the noise parameters.

The integral parameters of the flat interference component $A_{noise}^{int}$, $B_{noise}^{int}$ and $C_{noise}^{int}$ can be determined with a linear MLS estimation:

\begin{equation}\label{eq16}
	{\hat\Theta_{noise}}=(F^{T}F)^{-1}F^{T}{\tilde U_{noise}},
\end{equation}

where

\begin{equation}\label{eq17}
	F^{T}=\left\|
	\begin{array}{l l l l c}
		x_{1}&...&xi&...&x_{N_{IPSnoise}}\\
		y_{1}&...&y_{i}&...&y_{N_{IPSnoise}}\\
		1&...&1&...&1
	\end{array}
	\right\|
\end{equation}

where $x_{j}$ and $y_{j}$ are the angular coordinates of the $j^{th}$ pixel, which is used to estimate parameters of the flat  interference component; $N_{IPS_{noise}}$ is the number of AIFP pixels, which do not contain the object image.

To obtain the integral parameters (Eq. 16 - Eq. 17), only the AIFP pixels not belonging to the areas with the object images ($N_{IPSnoise}\leq(N_{IPS}-N_{IPSs})$) should be used. To exclude the influence of the anomalous emissions of the brightness in the pixels, we apply two iterations of MLS. In the second iteration, we use only those pixels, for which the value of the obtained frequency $\nu_{iktnoise}^{*}$ satisfies the following condition:

\begin{equation}\label{eq18}
	|\nu_{iktnoise}^{*}-{\hat\nu_{iktnoise}^{*}}|\leqslant\nonumber\\
\end{equation}

\begin{equation}\label{eq18}
	\leqslant
	k_{noise}\sqrt{\frac{\sum_{i,k}^{N_{IPS_{noise}}}(\nu_{iktnoise}^{*}-{\hat\nu_{iktnoise}^{*}})^{2}}{N_{IPS_{noise}}}},
\end{equation}

where $k_{noise}$ is the threshold coefficient for removing the pixels, which do not satisfy this condition, for example, $k_{noise}$=3;
$\sqrt{\frac{(\sum_{i,k}^{N_{IPSnoise}})(\nu_{iktnoise}^{*}-{\hat\nu_{iktnoise}^{*}})^{2}}{N_{IPSnoise}}}$
are the standard deviations of the flat interference component; and ${\hat\nu_{iktnoise}^{*}}={\hat A_{noise}^{int}}x_{i_{t}}+ {\hat B_{noise}^{int}}y_{k_{t}}{\hat C_{noise}^{int}}$ is the smoothed estimation of the measured frequency of the $ik^{th}$ pixel that is a part of the $N_{IPSnoise}$ value given in Eq. 16.

The number of pixels $N_{IPSnoise}$ determining the set ${\tilde U_{noise}}$ is reduced on its quantity, for which the condition (Eq. 18) is not satisfying. The process is repeated until one of the following conditions is completed: 1) the module of difference of the two related values of standard deviations becomes less than a certain value; 2) the number of pixels $N_{IPSnoise}$ becomes less than a given number; 3) the number of iterations exceeds a predetermined limit.

The next step is as follows: the obtained values of the integral parameters of the flat interference component are subtracted from the object signal of the given AIFP:

\begin{equation}\label{eq19}
	\nu_{ikts}^{*}=\nu_{ikts}^{*}-({\hat A_{noise}^{int}}x_{it_{s}}+{\hat B_{noise}^{int}}y_{kt_{s}}+{\hat C_{noise}^{int})},
\end{equation}

where $\nu_{ikts}^{*}$ is the measured frequency of the photons hitting the $ik^{th}$ pixel from the AIFP area; $x_{its}$ and $y_{kts}$ are the angular coordinates of the $ik^{th}$ pixel from the AIFP area.

The AIFP area for the calculation of the flat interference component of the object image is set by the operator (the default $ 31\times31 $ pixels). If the size of the object image is larger for this area, then the size of the area for the flat interference component is taken as the size of the object image multiplied by two (it is also set by the operator). As for the area of frame for fitting, we note that the fitting is carried out on the pixels that belong to the object image. Determination of these pixels is conducted through the delineation procedure, i.e. the fitting is carried out not for the fixed area but for the area which depends on the size of the object image.

The principal stages of the object image processing in the CoLiTec software are demonstrated at the pipeline in Fig. 1. They include: a) exclusion of stationary objects; b) detection of moving objects; c) analysis of moving objects, where A1, A2 and  A3 outline the coordinate deviations of moving objects from their trajectory. If it is necessary, the measurements can be stacked, but only un-stacked frames are statistically processed. Image service files (flat - bias - dark) can be used during the calibration process, but we offer an alignment frame option, which is commonly used. For example, to align frames, we used a high-pass Fourier filter in the earlier versions of the CoLiTec software. Currently we are using a median filtering, which has practically the same quality but it is substantially faster. To select the moving objects, the measurements (blobs) are formed in all the selected object images. After this, frames should be identified with each other and coordinates of all the measurements should be led to the basic frame.

The algorithm of the proposed method, which could be helpful to implement it, works as follows:

1.  To form a studied square area of infraframe processing (AIFP) with a side of $l$ pixels ($N_{IPS_{s}}$=$s^2$ ) and a square region of the presupposed existence of images of celestial objects with a side of $s$ pixels ($s<<l $ ). The centres of these regions are the local maxima of the image of the object (asteroid, comet, etc.) discovered previously.

2. To conduct a multipass MLS-estimation of parameters of the interface noise component according  to Eqs. 15 and 17. Wherein, only those pixels are processed at the next MLS-estimation pass, for which the value of the relative frequency $\nu^{*}_{ikt}$ (Eq. 1) satisfies the condition of Eq. 18. This do-while loop is repeated as long as the module of difference of the two standard deviation values obtained sequentially becomes smaller than a predetermined value.

3. To exclude the noise photons from the potentials of pixel in a region of the presupposed existence of images of celestial objects according to Eq. 19 (MLS-estimations of parameters of the interface noise component).

4. To estimate the coordinates of the object position according to Eq. 19 on the digital image, from which the noise photons were excluded earlier. Initial values of the coordinates of object position should be equal to the coordinates of the central local maximum of the AIFP image (AIFP centre):

4.1.  to calculate the coefficients of splitting $\lambda_{ikt}$  according to Eq. 10;

4.2. to estimate the coordinates of object position on the CCD frame according to Eq. 9;

4.3. to get the standard deviation estimate of the  photons hitting from the object according to Eq. 11;

4.4. to determine values of the weights of the discrete mix of the probability distributions according to Eq. 12;

4.5. to compare the actual value and value obtained in the previous step. If the difference between them is greater than a predetermined value, the actual value goes to the item 4.1 as estimates of the previous step; otherwise, it goes to the algorithm output as a result of its work.

The important step of the automated pipeline is also the creation of a catalogue of stationary objects (blobs) in the frame series. All the objects of this catalogue should be excluded in next stages of the processing but a few objects should be left as reference stars for astrometry.
Thus, the measurements, which are absent in the catalogue of stationary objects, have been used to select the moving objects. With this aim, the original method of collecting light (collection light technology) is provided.  Candidates to the small Solar system bodies (e.g. asteroids) selected by this method should have about the same brightness on different frames, and their coordinates should not deviate significantly from their average trajectory. In case of a partial occultation of a star by an asteroid, more often they stand out as two different celestial objects, wherein the position accuracies of the star and asteroid are falling.
In general, the detection of an asteroid is provided at the step of blobs, while the images themselves are not involved into processing (in this course, the blobs are formed according to the object images).

\section{Results and discussion}
The proposed method provides accurate estimation of asteroid coordinates on a set of the CCD frames and is the basis of the CoLiTec software, which was installed at several observatories of the world since March 1, 2009 (see, also, Fig.1).

\begin{figure*}
\begin{center}
\includegraphics[scale=0.32]{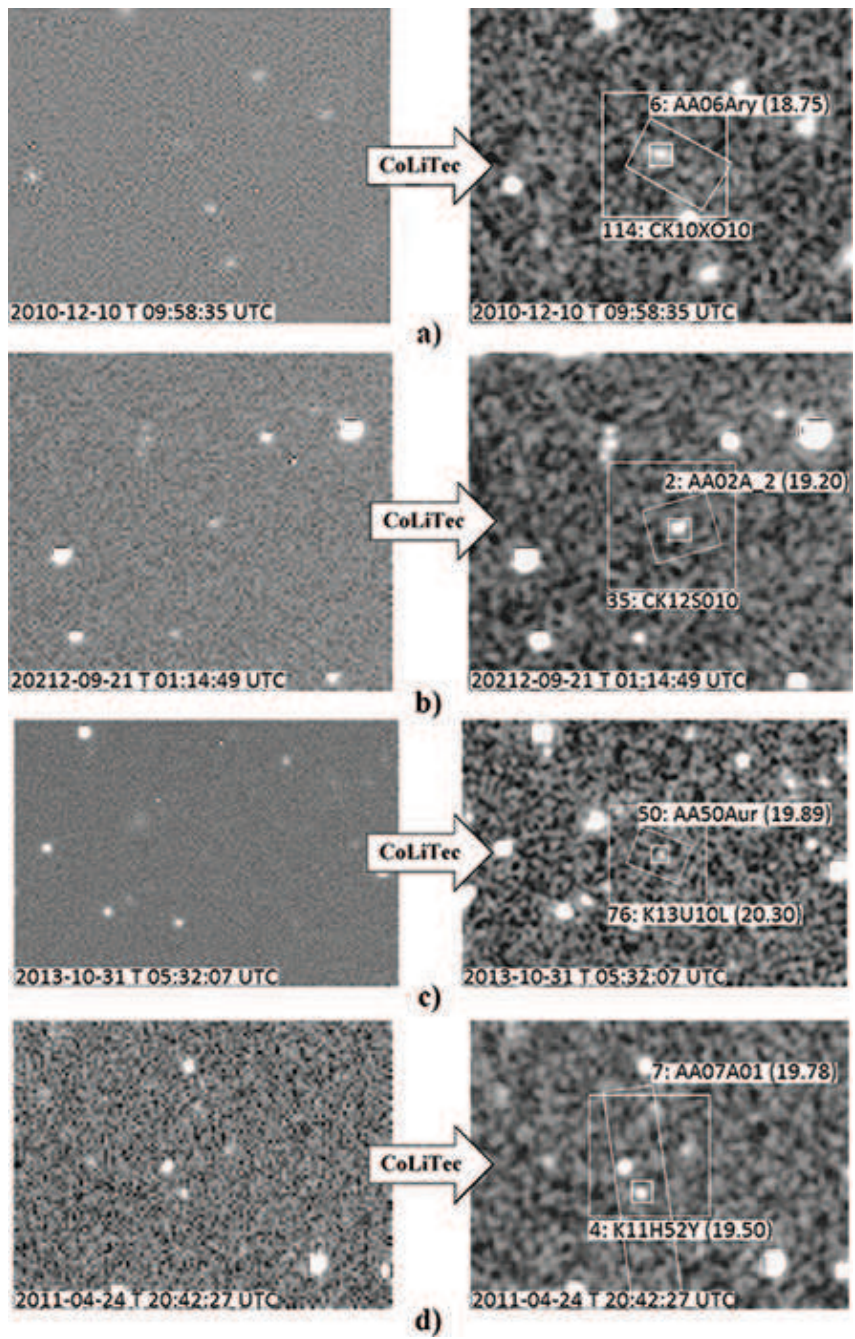}
\caption{The most known small Solar System bodies, which were detected and discovered
	with the CoLiTec software based on the subpixel Gaussian model:	
	a) the long-period Comet C/2010 X1 (Elenin), having an image size of 5 pixels, has been shifted by 7 pixels on a set of four CCD frames;
	b) the Comet C/2012 S1 (ISON), having an image size of 5 pixels, has been shifted by 3 pixels on a set of four CCD frames;	
	c) Centaur object 2013 UL10, having an image size of 5-6 pixels, has been shifted by 2 pixels;
	d) the K11H52Y asteroid, having an image size of 5 pixels, has been shifted by 45 pixels on a set of three CCD frames (it is described as unusual in the MPC circular  www.minorplanetcenter.net/mpec/K11/K11J02.html)}
\label{fig:discoveries}
\end{center}
\end{figure*}

In April, 2010, the CoLiTec was installed at the Andrushivka Astronomical Observatory (A50), Ukraine, and already in May, 2010, two asteroids were discovered (it was the first discovery of asteroids in the automatic mode at the observatories of the CIS countries).
On November 27, 2010, this software was installed at the ISON-NM (H15, New Mexico, USA), and on December 10, 2010, the comet C/2010 X1  was discovered by L. Elenin. In July, 2012, the CoLiTec software was installed at the ISON-Kislovodsk (D00, Russian Federation),
and on September 21, 2012, the comet  C/2012 S1(ISON) was discovered by V. Nevsky and A. Novichonok, the amateurs of astronomy. The digital images of discoveries of these comets are given in Fig.~\ref{fig:discoveries}a and Fig.~\ref{fig:discoveries}b, respectively.
As we mentioned in the Introduction, in total 4 comets and more than 1500 asteroids have been discovered with this software since 2010 (two examples of the asteroid discoveries are demonstrated in Fig.~\ref{fig:discoveries}c and Fig.~\ref{fig:discoveries}d, where the latter object discovered at the Andrushivka Astronomical Observatory on April 24, 2011, is marked as unusual K11H52Y asteroid in the MPC circular).

Since this method was already being operated during the asteroid observational surveys, we are able to provide a comparative analysis of the statistic parameters of accuracy estimations by this method and other results of thirty observatories that are most productive in terms of the number of asteroid observations in 2011-2013 (Tables 1-3). The observatories, which work with only one object in the centre of the CCD frame during the calm and suitable phase of the Moon, are excluded from our analysis. The observatories, where the CoLiTec software based on this method was installed, work during gusts and when it is calm. Moreover, the method is adaptive in such a way that there is no problem in automatic processing the object images on the CCD frames in the centre and on its edges as well as to provide measurements of many objects on the single frame.

Therefore,  in the years 2011-2013, such observatories as the ISON-NM Observatory (H15) (\citet{Elenin2012}, \citet{Elenin2013b}), Andrushivka Astronomical Observatory (A50) (\citet{Ivash2011}. \citet{Ivash2012}. \citet{Ivash2013}), and ISON-Kislovodsk Observatory (D00) acted as the users of the CoLiTec software (\citet{Elenin2014}). In the ranking of the most productive observatories worldwide in 2012, based on the number of discoveries of the small Solar System bodies, the users of the CoLiTec software had the 3rd, 13th, and 22nd places, respectively. As for the final report of 2011-2012, the ISON-NM Observatory (H15) holds the 7th place both by the number of measurements and priority of discoveries.

In Table 1 to 3, the total numbers of measurements and objects (Column 3) as well as of the asteroid discoveries (Column 4) are given according to the circulars of 2011, 2012, and 2013 of the Minor Planet Center (MPC), respectively (\citet{MPC}). The statistic parameters of these measurements are taken from the MPC site. The following parameters are also indicated in the tables for each observatory:
diameter $D$ of the primary mirror of the telescope, in meters (Column 5);
scale $S_{pix}$ of the pixel image, in arc seconds (Column 6);
the average residuals ${\bar\Delta_{\alpha}}$ and ${\bar\Delta_{\delta}}$ of object positions in the right ascension $\alpha$ and declination $\delta$ at a predetermined time (Column 7);
standard deviation estimations $\sigma_{\alpha}$ and $\sigma_{\delta}$ of object positions in the right ascension $\alpha$ and declination $\delta$ at a predetermined time (Column 8);
standard deviation estimations $\sigma^{\prime\prime}$ of object position (Eq. 20), in arc seconds (Column 9);
standard deviation estimations $\sigma_{pix}$ of object position (Eq. 21), in pixels (Column 10);
module of the Average Residuals of object position Measurements (Eq. 22), ARM (Column 11).
To calculate some aforementioned parameters, we applied the formula as follows:

\begin{equation}\label{eq20}
	\sigma^{\prime\prime}=0,5(\sigma_{\alpha}+\sigma_{\delta});
\end{equation}
\begin{equation}\label{eq21}
	\sigma_{pix}=\frac{\sigma^{\prime\prime}}{S_{pix}};
\end{equation}
\begin{equation}\label{eq22}
	ARM=\sqrt{({\bar\Delta_{\alpha}})^{2}+({\bar\Delta_{\delta}})^{2}}.
\end{equation}

\begin{table*}
	\caption{The accuracy parameter of thirty observatories that were the most productive in the number of asteroids measurements in 2011 according to the MPC data.}
	\centering
	\begin{tabular}{ccccccccccc}
		\hline
		(1)&(2)&(3)&(4)&(5)&(6)&(7)&(8)&(9)&(10)&(11)\\
		N&
		Observatory code&
		Measurements, objects&
		Discoveries&
		D, m&
		$S_{pix}$&
		R.A.($\bar\Delta_{\alpha}/\sigma_{\alpha})$&
		Decl.($\bar\Delta_{\delta}/\sigma_{\delta})$&
		$\sigma^{\prime\prime}$&
		$\sigma_{pix}$&
		ARM\\
		\hline
		1  & G96 & 2106367, 382737   & 21770      &1.50 &    1.00       &-0.01 +/- 0.32 & -0.04 +/- 0.28 & 0.300   &   0.30   & 0.041\\
		2  & 704 & 1956368, 279129   & 495    & 1.00 & 2.20  & 0.25 +/- 0.66  & 0.43 +/- 0.64 & 0.650      &   0.29    &      0.497\\
		3  & F51 & 1557902, 351923   & 13628      & 1.80 &  0.30       & 0.05 +/- 0.16 & 0.06 +/- 0.17 & 0.165 & 0.55 & 0.078\\
		4  & 703 & 1512387, 259412   & 2995       & 0.68&  2.60       &-0.21 +/- 0.67 & 0.17 +/- 0.68 & 0.675 & 0.25 & 0.270\\
		5  & 691 & 811571, 154495    & 8356       & 0.90 &  1.00       &-0.16 +/- 0.33 & 0.10 +/- 0.30 & 0.315 & 0.315 & 0.189\\
		6  & E12 & 219903, 52808     & 327        & 0.50 &  1.80$^1$      &-0.04 +/- 0.49 & 0.32 +/- 0.48 & 0.485 & 0.26 & 0.322\\
		7  & 645 & 208656, 45961     & 7          &  2.50   &    0.39        &           &           &       &      &       \\
		8  & D29 & 185303, 43414     & 318        & 1.04&  1.70       &           &           &       &      &       \\
		9  & C51 & 162900, 15412     & 23         &  0.40   &     2.75       & 0.06 +/- 0.57& -0.03 +/- 0.65 & 0.610  &   0.22   & 0.067\\
		\textbf{10} &  \textbf{H15}  & \textbf{154970, 37495}      & \textbf{768}         & \textbf{0.45} & \textbf{2.00}          &\textbf{-0.03 +/- 0.49} & \textbf{0.06 +/- 0.54} & \textbf{0.515} & \textbf{0.25} & \textbf{0.067}\\
		11 & 106 & 75340, 18093      & 73         & 0.60 &  2.00         & 0.04 +/- 0.36& -0.11 +/- 0.35 & 0.355 & 0.17 & 0.117\\
		12 & 291 & 70355, 19028      & 646        & 1.80 &  0.60       &-0.13 +/- 0.36 & 0.15 +/- 0.27 & 0.315 & 0.52 & 0.191\\
		13 & J75 & 48469, 13209      & 561        & 0.45&  1.47          &-0.04 +/- 0.42& -0.14 +/- 0.40 & 0.410  & 0.28     & 0.146\\
		14 & 644 & 34164, 6255       & 954        &  1.20   &      1.00      &           &           &       &      &       \\
		\textbf{15} & \textbf{A50}  & \textbf{33386, 9755}       & \textbf{72}          & \textbf{0.60} & \textbf{2.06}       &\textbf{-0.03 +/- 0.51} & \textbf{0.05 +/- 0.51} & \textbf{0.510}  & \textbf{0.24} & \textbf{0.058}\\
		16 & 926 & 28578, 8460       & 171        & 0.81, 0.41 & 0.87 & 0.15 +/- 0.38 & 0.27 +/- 0.39 & 0.385 & 0.44 & 0.309\\
		17 & 461 & 28038, 6281       & 782        & 0.60,  1.02& 1.10  &-0.03 +/- 0.27 & 0.14 +/- 0.27 & 0.270  & 0.24 & 0.143\\
		18 & A14 & 24354, 6448       & 115        & 0.50 &           & 0.08 +/- 0.41& -0.06 +/- 0.36 & 0.385 &      & 0.100\\
		19 & J04 & 23322, 6460       & 188        & 1.00  &  0.62$^2$    & 0.16 +/- 0.29 & 0.24 +/- 0.30 & 0.295 & 0.47 & 0.288\\
		20 & A77 & 21677, 5423       & 318        &   0.50   &           & 0.027 +/- 0.63& 0.22 +/- 0.50 & 0.565 &      & 0.348\\
		\multicolumn{11}{l}{$^1$\citet{Mahabal1}, $^2$\citet{Li2}, $^3$\citet{Ory}}\\
		\multicolumn{11}{l}{(fn) Highlighted columns are CoLiTec users.}\\
		\hline
	\end{tabular}
	\label{tab:t-agn}
\end{table*}

\begin{table*}
	\caption{The accuracy parameter of thirty observatories that were the most productive in the number of asteroids measurements in 2012 according to the MPC data.}
	\centering
	\begin{tabular}{ccccccccccc}
		\hline
		(1)&(2)&(3)&(4)&(5)&(6)&(7)&(8)&(9)&(10)&(11)\\
		N&
		Observatory code&
		Measurements,objects&
		Discoveries&
		D, m&
		$S_{pix}$&
		R.A.($\bar\Delta_{\alpha}/\sigma_{\alpha})$&
		Decl.($\bar\Delta_{\delta}/\sigma_{\delta})$&
		$\sigma^{\prime\prime}$&
		$\sigma_{pix}$&
		ARM\\
		\hline
		1 & G96 & 2080033, 384204 & 17676 & 1.50  &  1.00   & 0.20 +/- 0.33    & 0.20 +/- 0.28  & 0.305     &  0.305    & 0.028\\
		2 & F51 & 1948353, 467091 & 13785 & 1.80   & 0.30 & 0.07 +/- 0.15  & 0.04 +/- 0.17  & 0.16      & 0.53 & 0.081\\
		3 & 703 & 1723293, 282864 & 2278  & 0.68  & 2.60 &-0.22 +/- 0.65  & 0.07 +/- 0.62  & 0.635     & 0.24 & 0.231\\
		4 & 704 & 1681504, 262209& 224   & 1.00   & 2.20 & 0.26 +/- 0.67  & 0.43 +/- 0.64  & 0.655     & 0.29 & 0.502\\
		5 & 691 & 896972, 163714  & 7600  & 0.90   & 1.00 &-0.16 +/- 0.32  & 0.10 +/- 0.29  & 0.305     & 0.27 & 0.189\\
		6 & E12 & 259295, 62621   & 430   & 0.50   & 1.80$^1$ &-0.01 +/- 0.51  & 0.29 +/- 0.50  & 0.505     & 0.28 & 0.290\\
		7 & J43 & 102641, 22682   & 531   & 0.50$^2$   & 1.20$^2$ & 0.19 +/- 0.48  & 0.05 +/- 0.40  & 0.44      & 0.36 & 0.196\\
		8 & 926 & 100161, 29986 & 454 & 0.81, 0.41 & 0.87 & 0.02 +/- 0.37 & 0.05 +/- 0.35 & 0.36 & 0.41 & 0.54\\
		\textbf{9} & \textbf{H15}  & \textbf{97878, 24170}    & \textbf{338}    & \textbf{0.45}   & \textbf{2.00} & \textbf{-0.06 +/- 0.50} & \textbf{-0.01 +/- 0.53}  & \textbf{0.515}     & \textbf{0.25} & \textbf{0.061}\\
		10& 106 & 72192, 17451    & 120   & 0.60   & 2.00 & 0.04 +/- 0.36 & -0.12 +/- 0.34  & 0.35      & 0.17 & 0.126\\
		11& A14 & 57243, 16239    & 159   & 0.50  &     & 0.06 +/- 0.37 & -0.02 +/- 0.32  & 0.345     &      & 0.063\\
		12& J04 & 43209, 10708    & 513   & 1.00   & 0.62$^3$ & 0.21 +/- 0.28  & 0.20 +/- 0.27  & 0.275     & 0.44 & 0.29\\
		\textbf{13} &\textbf{D00}  & \textbf{31494, 7403}     &\textbf{61}     & \textbf{0.40}  &\textbf{2.06} & \textbf{0,00 +/- 0,57} & \textbf{-0,06 +/- 0,41} & \textbf{0,49}       & \textbf{0,23} & \textbf{0.06}\\
		14& 291 & 24272, 6224     & 28    & 1.80   & 0.60 & 0.07 +/- 0.33  & 0.13 +/- 0.28 & 0.305      & 0.50 & 0.148\\
		15& 461 & 23847, 5615   & 170 & 0.60, 1.02 & 1.10 & 0.00 +/- 0.27   & 0.15 +/- 0.27 & 0.27       & 0.24 & 0.15\\
		16& 644 & 22714, 4486     & 332   & 1.20$^4$   & 1.00$^4$  &            &           &            &      &       \\
		17& H21 & 22672, 3870 & 181 & 0.61, 0.81, 0.76 & 0.80$^2$  & 0.03 +/- 0.34 & 0.01 +/- 0.36 & 0.35       & 0.43 & 0.032\\
		18& I41 & 21245, 2392     & 1790  & 1.20$^5$    & 1.01$^5$ & 0.11 +/- 0.23 & -0.03 +/- 0.23 & 0.23       & 0.22 & 0.114\\
		19& A24 & 18940, 2412     & 0     & 0.36  & 1.40 & 0.14 +/- 0.37  & 0.24 +/- 0.33 & 0.35       & 0.25 & 0.278\\
		\textbf{22} & \textbf{A50} &\textbf{11559, 3725}      &\textbf{13}     & \textbf{0.60}   &\textbf{2.07} &\textbf{0.25 +/- 0.50}  &\textbf{-0.04 +/- 0.46}  & \textbf{0.48}       & \textbf{0.23} & \textbf{0.253}\\
		\multicolumn{11}{l}{$^1$\citet{Mahabal1}, $^2$\citet{Ory}, $^3$\citet{Abreu}, $^4$\citet{NEAT-PALOMAR}, $^5$\citet{Waszczak}}\\
		\multicolumn{11}{l}{(fn) Highlighted columns are CoLiTec users.}\\
		\hline
	\end{tabular}
	\label{tab:t-agn}
\end{table*}

\begin{table*}
	\caption{The accuracy parameter of thirty observatories that were the most productive in the number of asteroids measurements in 2013 according to the MPC data.}
	\centering
	\begin{tabular}{ccccccccccc}
		\hline
		(1)&(2)&(3)&(4)&(5)&(6)&(7)&(8)&(9)&(10)&(11)\\
		N&
		Observatory code&
		Measurements, objects&
		Discoveries&
		D, m&
		$S_{pix}$&
		R.A.($\bar\Delta_{\alpha}/\sigma_{\alpha})$&
		Decl.($\bar\Delta_{\delta}/\sigma_{\delta})$&
		$\sigma^{\prime\prime}$&
		$\sigma_{pix}$&
		ARM\\
		\hline
		1 & F51 & 2279609, 506894 & 14168 & 1.80  & 0.30& 0.13 +/- 0.07    & 0.06 +/- 0.14   & 0.135    & 0.45    & 0.092\\
		2 & G96 & 1950642, 343808 & 11908 & 1.50 & 1.00 & 0.04 +/- 0.32  & 0.05 +/- 0.28   & 0.300      & 0.30    & 0.064\\
		3 & 703 & 1844330, 289086 & 1494  & 0.68 & 2.60  & -0.14 +/- 0.66 & 0.22 +/- 0.64   & 0.650     & 0.25    & 0.260\\
		4 & 691 & 742001,  139225 & 5594  & 0.90  & 1.10  & -0.16 +/- 0.31 & 0.12 +/- 0.30   & 0.315    & 0.28    & 0.200\\
		5 & D29 & 551094, 136964  & 262   & 1.04 & 1.70  & 0.03 +/- 0.53 & -0.04 +/- 0.49   & 0.510     & 0.30    & 0.050\\
		6 & I41 & 440712, 52579   & 2270  & 1.20$^1$ & 1.01$^1$&  0.06 +/- 0.18 & 0.02 +/- 0.17   & 0.175    & 0.17    & 0.063\\
		7 & E12 & 229747, 48026   & 204   & 0.50  & 1.80$^2$  &-0.02 +/- 0.50 & 0.28 +/- 0.46   & 0.480     & 0.26    & 0.280\\
		8 & 926 & 179570, 53662 & 750 & 0.81, 0.41& 0.87 & 0.15 +/- 0.39 & 0.10 +/- 0.36 & 0.375   & 0.43 & 0.180\\
		9 & J43 & 151983, 27006   & 1006  & 0.50$^3$  & 1.20$^3$  & 0.11 +/- 0.31 & -0.03 +/- 0.29   & 0.300      & 0.25    & 0.114\\
		10& W84 & 110213, 8518    & 4160  & 40$^{3,4}$    & 0.27$^4$ & 0.13 +/- 0.13 & 0.14 +/- 0.13    & 0.130     & 0.48    & 0.191 \\
		\textbf{11} &\textbf{H15}  &\textbf{107989, 25282}    &\textbf{156}    & \textbf{0.40} & \textbf{2.00}  & \textbf{0.09 +/- 0.62} & \textbf{0.02 +/- 0.60}    & \textbf{0.610}     & \textbf{0.305}   & \textbf{0.092}\\
		12& 704 & 81054, 17833    & 4     & 1.00  & 2.20  & 0.29 +/- 0.64 & 0.38 +/- 0.63    & 0.635    & 0.28    & 0.478\\
		13& J04 & 58307, 14670    & 576   & 1.00  & 0.62$^5$ & 0.25 +/- 0.30 & 0.23 +/- 0.28    & 0.290     & 0.46    & 0.340\\
		\textbf{14} &\textbf{D00}  & \textbf{44658, 10850}    &\textbf{34}     & \textbf{0.40} &\textbf{2.06}  &\textbf{0.01 +/- 0.72} & \textbf{-0.12 +/- 0.54}    &\textbf{0.630}      &\textbf{0.305}    &\textbf{0.120}\\
		15& G32 & 36416, 4654     & 654   & 0.40  & 1.13 & 0.03 +/- 0.35 & 0.05 +/- 0.32    & 0.335    & 0.29    & 0.058\\
		16& 106 & 18601, 4502     & 67    & 0.60  & 2.00  & 0.04 +/- 0.39 & -0.05 +/- 0.37   & 0.370     & 0.19    & 0.064\\
		17& H21 & 16924, 2994 & 60 & 0.61, 0.81, 0.76 & 0.80$^6$ & 0.04 +/- 0.33 & -0.04 +/- 0.31 & 0.320& 0.4 & 0.002\\
		18& 461 & 15688, 3787 & 110  & 0.60, 1.02 & 1.10  & -0.02 +/- 0.24 & 0.17 +/- 0.27    & 0.255    & 0.23    & 0.171\\
		19& 644 & 15221, 3317     & 63    & 1.20$^7$   & 1.00$^7$     &           &              &          &         &\\
		20& 291 & 15197, 4002     & 1     & 1.80  & 0.60  & 0.02 +/- 0.35 & 0.14 +/- 0.32    & 0.335    & 0.55    & 0.141\\
		\multicolumn{11}{l}{$^1$\citet{Waszczak}, $^2$\citet{Mahabal1}, $^3$\citet{Ory}, $^4$\citet{Honscheid}, $^5$\citet{Abreu}, $^6$\citet{Li2}, $^7$\citet{NEAT-PALOMAR}}\\
		\multicolumn{11}{l}{(fn) Highlighted columns are CoLiTec users.}\\
		\hline
	\end{tabular}
	\label{tab:t-agn}
\end{table*}
	
	\begin{figure*}
		\begin{center}
			\includegraphics[scale=0.18]{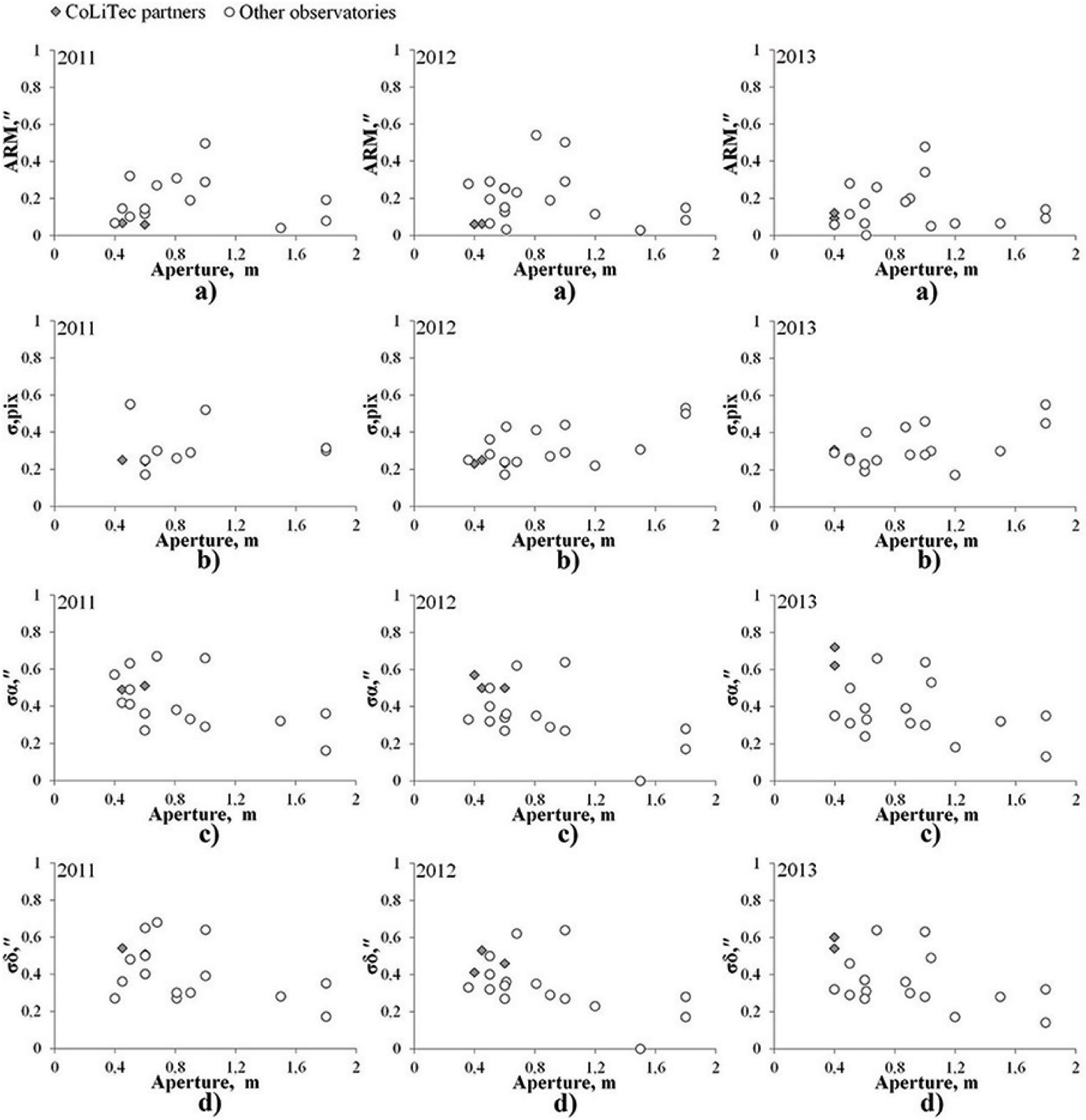}
			\caption{The comparative analysis of the accuracy parameters of object position for the most productive observatories in the number of asteroids measurements in 2011 (left), 2012 (middle), and 2013 (right) by:
				a) module of the average residuals of object position measurements;
				b) standard deviation estimations $\sigma_{pix}$ of object position, in pixels;
				c) standard deviation estimations of object position in the right ascension, $\sigma_{\alpha}$, in arcseconds;
				d) standard deviation estimations of object position in the declination, $\sigma_{\delta}$, in arcseconds. }
			\label{fig:statistics}
		\end{center}
	\end{figure*}

The data analysis related to the accuracy parameters of object position for the most productive observatories in the number of asteroids measurements in 2011-2013 is illustrated in Figure 3.

The observatory-partners of the CoLiTec program hold leading positions in their class of telescopes as related to the parameter of module of the average residuals of measurements (Figures 3a).  In 2011 and 2012, this parameter for H15 (\citet{Elenin2012}) and A50 observatories was equal to 0.06$^{\prime\prime}$. At the same time, these observatories were not in the list of the best ones as related to the parameters of standard deviation estimations of object position (in arc seconds) (see, Figures 3c and 3d).
In 2011, the values of standard deviation estimations $\sigma^{\prime\prime}$ of object position for the mentioned observatories were equal to 0.515$^{\prime\prime}$(H15) and 0.51$^{\prime\prime}$ (A50). In 2012, these values were equal to 0.515$^{\prime\prime}$ (H15), 0.49$^{\prime\prime}$ (D00), and 0.48$^{\prime\prime}$ (A50). The reason for such a deterioration of the results, in addition to the size of the aperture of a telescope, is the pixel scale of the used CCD matrix. To take this factor into account, the observatory-partners of the CoLiTec program decided to consider the parameter of standard deviation estimations $\sigma_{pix}$ of object position, in pixels, for accurate estimation of asteroid coordinates on the CCD frame as a principal one during observations.

The parameter of standard deviation estimations $\sigma_{pix}$ of the object position in the pixels on the CCD frame (Figures 3b; Column 10 in Tables 1-3) is used to characterise namely an efficiency of the mathematical method, which is applied for coordinate measurements. In other words, it allows the observer to be disengaged from parameters of the used CCD matrix and other devices.
According to this parameter, the observatory-partners of the CoLiTec program have one of the best results among the observatories in their class of telescopes (small aperture). In 2011 (2012), this parameter was equal to 0.25 (0.25) pixel and 0.24 (0.23) pixel for observatories H15 and A50, respectively, as well as was equal to 0.23 pixel for D00 observatory in 2012.
In 2013, the accuracy of measurements of the observatories (CoLiTec partners) fell approximately 20\% for all the mentioned parameters due to an error in software that was immediately fixed and corrected completely. As a result, the observatory-partners of the CoLiTec program returned to their previous positions as for the indexes of measurement accuracy (see, in detail, the current MPC site).
It is important to emphasise that standard deviation $\sigma_{pix}$ of object position (Figures 3b) is an artificial parameter. It is not a strong objective because such factors as the exposure time, the telescope optical scheme, the height above sea level, and many others are not taken into account in it. For example, according to this parameter, the Pan-STARRS 1 observatory (F51) had lost positions, although it had the best astrometry accuracy among all the asteroid surveys. We may suggest that an implementation of our method, which is free from a possible loss of measurement information contained on the CCD frames, when using at this observatory most likely can yield the best results than that was used.

We also compared the CoLiTec software and the Astrometrica that is widely accepted among amateurs.

Both the softwares: realise such functions as
frame calibration;
support of astrometric and photometric catalogues of stars in the local and online modes;
record the coordinate information (WCS) in the FITS-frame title;
have interactive mode for the object position measurements;
have a magnifier tool;
have an automated search of moving objects;
have a mode for visual inspection of detected moving objects;
provide output of astrometric measurements in the MPC format;
transfer data mode from the interface to the MPC;
display the known and discovered objects on the CCD frame.
However, the CoLiTec does not realise such functions as
Track$ \& $Stack for adding the frame, and
identification of detected moving objects with a local database MPCORB.

At the same time, unlike the Astrometrica, the CoLiTec: realises such functions as
astrometric reduction of the CCD frames with large fields of view ($ 2^{\deg} $ or more);
separates application of the reference catalogues for astrometric and photometric reductions;
provides an automated search of moving faint objects (SNR ~ 2.5);
has automatic data processing;
saves results on the processing frames (the real detected objects; objects, which are rejected by the operator, etc.);
identifies the detected moving objects with the online database MPCORB (MPC);
identifies the stationary objects with database of variable stars (VSX) and galaxies (HyperLeda);
has a software modular design (the ability to connect the individual modules).
The CoLiTec software gives more accurate measurements of faint celestial objects as well as
contain a more reliable method for identifying the frames with a reference star catalogue
allowing to improve (sometimes significantly) an accuracy of the object position.

Along with the analysis of indexes of measurement accuracy (see, the current MPC site), we conducted a comparative analysis of the accuracy of both the softwares (CoLiTec vs. Astrometrica) after processing the same frame. We selected 19 series for each of the 4 frames. A preliminary analysis included 36 series. However, the rest of the series had no reliable identification with the used star catalogue. Also, we excluded frames with significant disruptions in the daily maintenance and frames taken at very high wind. All frames were obtained at the observatory ISON-NM (H15) with help of a 40-cm telescope Santel-400AN and CCD-matrix FLI ML09000-65 ($3056 \times 3056$ pixels, a pixel size of 12 microns) in the period from March 4, 2014 to March 30, 2014. The exposure time was 150 seconds.

We used only such object positions, the measurements of which are included in the MPCAT-OBS archive (\citet{MPCATOBS}). The measurements, however, were reprocessed with the CoLiTec software. To set the reference values of object positions on the time of measurements formation, we used the HORIZONS service (\citet{HORIZONS}). In total, we used 2002 measurements (measurements with CoLiTec were done on 253 more). The Astrometrica in several cases issued a sign about impossibility to guarantee a reliable measurement of object position (Centroid = -1). It is often associated with an attempt to measure the position of star trails or faint objects involved with a brighter star. More than half of them have an SNR not exceeding 3.5. The results of a comparative analysis are shown in Table 4. One can see that measurements with the Astrometrica at a low SNR have an RMS of 30-50\% larger than that of the CoLiTec (see, also Fig. 4 and 5). Mean deviations in measurements with the CoLiTec and the Astrometrica are the same in general, and so these data are not shown. The more detailed comparative analysis is given in our paper (\citet{Sav15}).

\begin{table}
	\caption{Comparative analysis of the softwares CoLiTec and Astrometrica as concerns with measurements of object positions for the numbered asteroids}
	\centering
	\begin{tabular}{ccc}
		\hline
		(1)&(2)&(3)\\
		Deviation&
		Astrometrica&
		CoLiTec\\
		\hline
		The average deviation of RA, $^{\prime\prime}$ &  0,11 &  0,11 \\
		The average deviation of DE, $^{\prime\prime}$ & -0,04 & -0,03 \\
		RMS deviation of RA, $^{\prime\prime}$         &  0,77 &  0,50 \\
		RMS deviation of RA, $^{\prime\prime}$         &  0,67 &  0,39 \\
		\hline
	\end{tabular}
	\label{tab:t-agn}
\end{table}

\begin{figure}
	\begin{center}
	\includegraphics[scale=0.29]{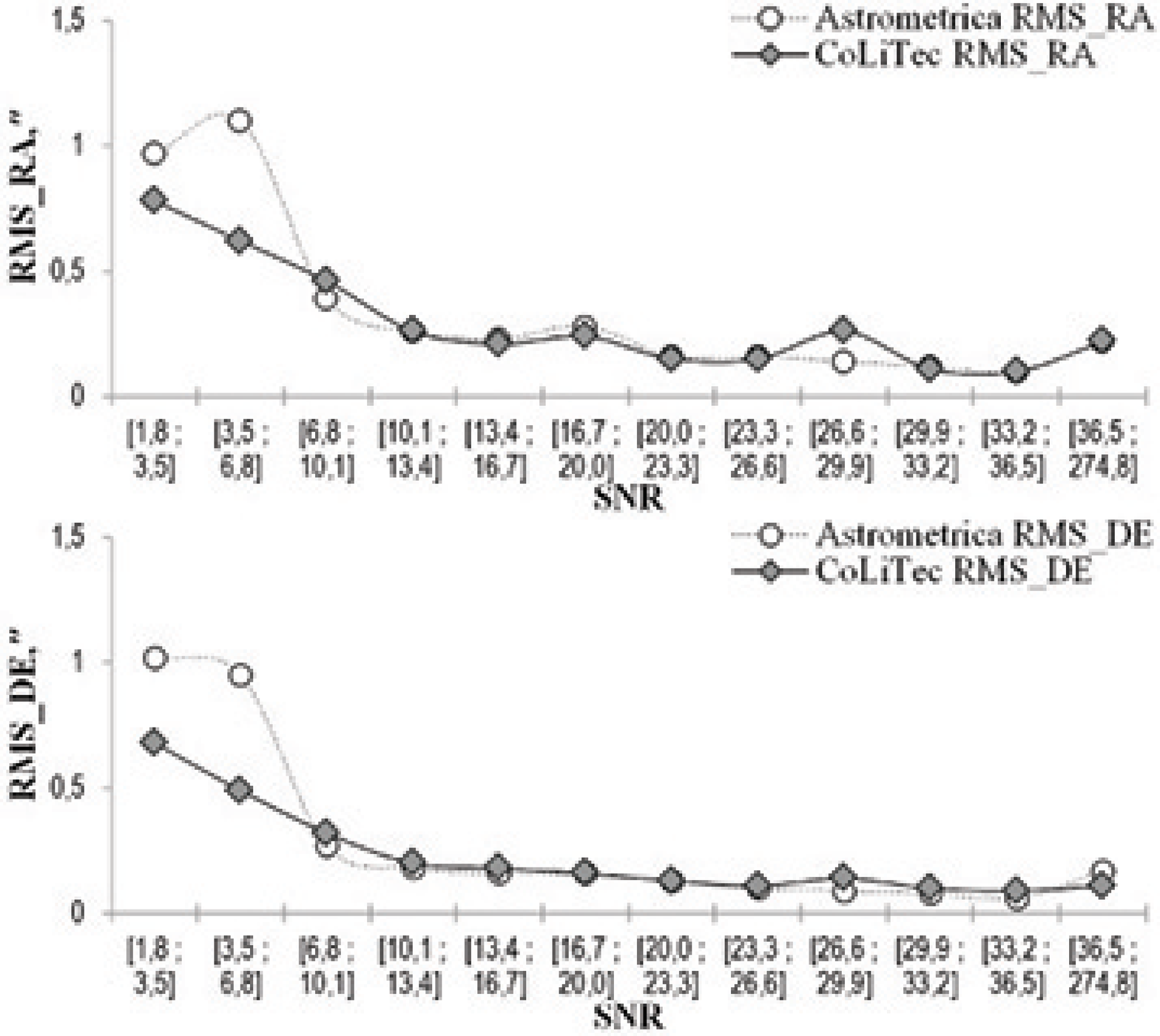}
		\caption{Distribution of deviations of the equatorial coordinates of object by SNR ranges (Astrometrica vs. CoLiTec)}
		\label{fig:SNR_RMS_RA_DE}
	\end{center}
\end{figure}

\begin{figure}
	\begin{center}
	\includegraphics[scale=0.29]{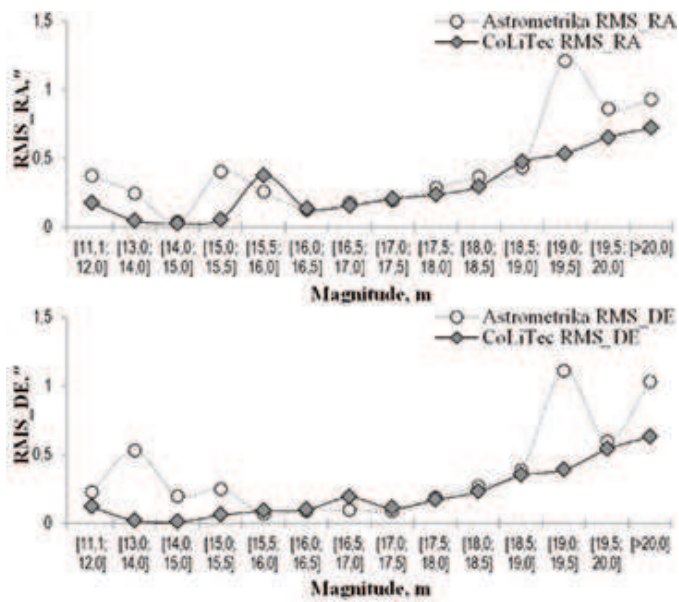}
		\caption{Distribution of deviations of the equatorial coordinates of object by magnitude ranges (Astrometrica vs. CoLiTec)}
		\label{fig:M_RMS_RA}
	\end{center}		
\end{figure}

\section{Conclusions}
A new iteration method of asteroid coordinate estimation on the digital image has been developed. The method operates by continuous parameters (asteroid coordinates) in a discrete observation space (the set of pixels potential of the CCD matrix).

High rates of the CoLiTec program during 2011-2012 as concerns with the accuracy of measurements have been obtained due to the use of the subpixel Gaussian model. This model of the object image takes into account a prior form of the object image and consequently it is adapted more easily to any forms of real image. In other words, nevertheless that a real coordinate distribution of  photons hitting the pixels on the CCD frame is not known, the form of this distribution is known a priori and its parameters can be estimated according to the real object image. At that time many other methods mentioned in Introduction consider by default that the density of hit photons inside the pixel is uniform.

The advantages of subpixel Gaussian model become more obvious for the fainter celestial objects. Moreover, the developed method has a high measurement accuracy as well as a low calculating complexity because
the maximum likelihood procedure is implemented to obtain the best fit instead of the least-squares method and Levenberg-Marquardt algorithm for the minimisation of the quadratic form.

The efficiency of the proposed method, including its advantages for accurately estimating asteroids coordinates, was confirmed during observations as the part of the  CoLiTec (Collection Light Technology) program for automatic discoveries of asteroids and comets on a set of the CCD frames. Efficiency is a crucial factor in the discovery of near-Earth asteroids (NEA) and potentially-hazardous
asteroids. Current asteroid surveys yield many images per night. It is no longer possible for the observer to quickly view these images in the blinking mode. It causes a serious difficulty for large-aperture wide-field telescopes, capturing up to several tens of asteroids in one image.
The CoLiTec software solves the problem of the frame processing for asteroid surveys in a real-time mode. We compared also our software and Astrometrica, which is widely used for detecting the newly objects. Limits of measurements of CoLiTec software are wider than of Astrometrica, but that the most valuable, this expansion comes into area of extremely small SNR allowing to search the fainter Solar System small bodies (measurements with the Astrometrica at a low SNR have an RMS of 30-50\% larger than that of the CoLiTec). As for the area of SNR$ > $7, the results of CoLiTec and Astrometrica are approximately identical. However, exactly the area of extremely small SNR is more promising for the discovery of new celestial objects.

The automatically detected small Solar System bodies are subject to follow-up visual confirmation. The CoLiTec software is in use for the automated detection of asteroids at the Andrushivka Astronomical Observatory, Ukraine (since 2010), at the Russian remote observatory ISON-NM (Mayhill, New Mexico, USA) since 2010, at the observatory ISON-Kislovodsk since 2012,
and at the ISON-Ussuriysk observatory since 2013 (see Tables 1-3). As the result, four comets (C/2010 X1 (Elenin) (\citet{Elenin2010}), P/2011 NO1(Elenin) (\citet{Elenin2011}, \citet{Elenin2013}), C/2012 S1 (ISON) (\citet{Nevski2012}), and P/2013 V3 (Nevski) (\citet{Nevski2013}) as well as more than 1500 small Solar System bodies (including five NEOs, 21 Trojans of Jupiter, and one Centaur object) have been discovered.

In 2014 the CoLiTec software was recommended to all members of the Gaia-FUN-SSO network (a network for Solar System transient Objects) for analysing observations as a tool to detect faint moving objects in frames. Information about CoLiTec with link to web-site has been posted on the Gaia-FUN-SSO Wiki (\url{https://www.imcce.fr/gaia-fun-sso/}).

The authors thank Dr. F. Velichko for his useful comments. We are grateful to the reviewer for his helpful remarks that improved this work. We express our gratitude to Mr. W. Thuillot, coordinator of the Gaia-FUN-SSO network, for the approval of CoLiTec as a well-adapted software to the Gaia-FUN-SSO conditions of observation. We also thank Dr. Ya. Yatskiv for his support of this work in frames of the Target Program of Space Science Research of the National Academy of Science of Ukraine (2012-2016) and the Ukrainian Virtual Observatory (\url{http://www.ukr-vo.org}). The CoLiTec program is available through \url{http://colitec.neoastrosoft.com/en} (one can access to the download package \url{http://www.neoastrosoft.com/download_en} and some instructions \url{http://www.neoastrosoft.com/documentation_en}).

\label{lastpage}
\end{document}